\documentclass[twoside,leqno]{article}

\usepackage[letterpaper]{geometry}

\usepackage{siamproceedings}

\usepackage[T1]{fontenc}
\usepackage{amsfonts,amssymb,siunitx,bbm,commath}
\usepackage{graphicx,tikz,pgfplots,booktabs}
\usetikzlibrary{ipe}
\usepackage{epstopdf}
\usepackage{enumitem}
\usepackage{algorithmic}
\ifpdf
  \DeclareGraphicsExtensions{.eps,.pdf,.png,.jpg}
\else
  \DeclareGraphicsExtensions{.eps}
\fi

\newsiamremark{remark}{Remark}
\newsiamremark{defn}{Definition}
\newsiamremark{hypothesis}{Hypothesis}
\crefname{hypothesis}{Hypothesis}{Hypotheses}
\newsiamthm{claim}{Claim}

\usepackage{amsopn}

\newcommand{\E}{\mathbb{E}}
\newcommand{\Z}{\mathbb{Z}}

\renewcommand{\P}{\mathbb{P}}
\newcommand{\R}{\mathbb{R}}
\newcommand{\eps}{\varepsilon}

\begin{document}

\newcommand\relatedversion{}

\title{\Large Dimension dependence of critical phenomena in long-range percolation}
    \author{Tom Hutchcroft\thanks{Division of Physics, Mathematics and Astronomy, California Institute of Technology  (\email{t.hutchcroft@caltech.edu}).}}
\date{}

\maketitle







\begin{abstract} 
Statistical mechanical systems at and near their points of phase transition are expected to exhibit rich, fractal-like behaviour that is independent of the small-scale details of the system but depends strongly on the dimension in which the model is defined. Moreover, many models are conjectured to have an \emph{upper critical} dimension with important quantitative and qualitative differences between critical behaviour at, above, and below the upper critical dimension. For models with \emph{long-range interactions}, one expects additional transitions between \emph{effectively long-range} and \emph{effectively short-range} regimes, with further marginal effects on the boundary of these two regimes, leading to (at least) eight qualitatively distinct forms of critical behaviour in total for each given model. We give a broad overview of these conjectures aimed at a general mathematical audience before surveying the significant recent progress that has been made towards understanding them in the context of long-range percolation.
\end{abstract}

\section{Universality and dimension dependence of critical phenomena.}
\label{sec:universality_and_dimension_dependence_of_critical_phenomena_}
The goal of this article is to survey a series of recent works concerning critical phenomena in \emph{long-range percolation} \cite{baumler2022isoperimetric,LRPpaper1,LRPpaper2,LRPpaper3,hutchcroft2024pointwise,hutchcroft2020power,hutchcroft2022sharp,hutchcrofthierarchical,hutchcroft2022critical}. We begin by giving a broad overview of phase transitions and critical phenomena in \Cref{sec:universality_and_dimension_dependence_of_critical_phenomena_} before specializing to percolation in \Cref{sec:critical_exponents_and_scaling_limits_for_bernoulli_percolation_the_dream_}, focussing on the transition from mean-field critical behaviour to low-dimensional critical behaviour at the upper critical dimension. Long-range models and their conjectured critical behaviours are introduced in \Cref{sec:long_range_models_}. Finally, in \Cref{sec:recent_progress_on_long_range_percolation} we explain how almost the entire heuristic theory of percolation critical phenomena we introduce in the first three sections can be verified for long-range percolation in the effectively long-range regime, including for models at, above, and below the upper critical dimension.


Let us start our discussion from a high vantage point.
Many large, complex systems undergo \emph{phase transitions}, meaning that the behaviour of the system changes in a stark, qualitative way as the parameter(s) governing the system at the microscopic level are varied continuously through a special value (or hypersurface of values). We are all familiar with the phase transitions that occur when e.g.\ water freezes or boils; other important examples of phase transitions from inside and outside the natural sciences  include iron's transition between \emph{ferromagnetic} and \emph{paramagnetic} behaviour at its \emph{Curie temperature} 
 \cite{spaldin2010magnetic}, the emergence of \emph{superconductivity} and \emph{superfluidity} in certain materials at very low temperature \cite{feynman1957superfluidity}, the phenomenon of \emph{herd immunity} in epidemiology~\cite{meyers2007contact}, and phase transitions in the average-case computational complexity of optimization problems \cite{ding2015proof}. In each case, understanding when, how, and why the system undergoes a phase transition is of central importance in both theory and practice. Moreover, the basic mathematical principles underlying the occurrence of such phase transitions have much in common across these diverse situations, and the study of phase transitions has come to be recognised as a deep wellspring of pure mathematics that is of interest beyond and complementary to its practical origins.


In addition to possessing phase transitions, many systems of interest also exhibit \emph{critical phenomena} at and near their points of phase transition: rich, fractal-like behaviours characterised by \emph{scale-invariance} and \emph{power laws} holding approximately at large scales. Within the probabilistic study of lattice models,
such critical phenomena are studied either via \emph{critical exponents}, which govern the power-law growth or decay of various quantities at or near criticality, or through \emph{scaling limits} (a.k.a.\ \emph{continuum limits}), in which one takes the limit of the ``entire system'' as the lattice spacing goes to zero, yielding a more complete picture of large-scale critical behaviour than provided by critical exponents alone. A prototypical example is the \emph{simple random walk} $(X_n)_{n\geq 0}$ on $\Z^d$, i.e., the random process on the $d$-dimensional hypercubic lattice $\Z^d$ that moves by $\pm 1$ in a random coordinate direction at each step, with different steps independent. The simple random walk has \emph{displacement exponent} $1/2$ in the sense that  $\|X_n\| $ is of order $n^{1/2}$ with high probability as $n\to \infty$ and, in dimension $d\geq 2$, has \emph{fractal dimension} $2$ in the sense that
$\{X_n:n\geq 1\}$ has intersection of order $r^{2\pm o(1)}$ with the box $[-r,r]^d$ with high probability for each $r\gg 1$. The simple random walk also
 has a well-defined scaling limit known as \emph{Brownian motion}  \cite{morters2010brownian}: a random continuous, nowhere-differentiable $\R^d$-valued function $(B_t)_{t\geq 0}$ such that
\[
  \left(\lambda^{-1/2} X_{\lfloor \lambda t \rfloor}\right)_{t\geq 0} \longrightarrow (B_t)_{t\geq 0}
\]
in distribution as the scale factor $\lambda$ tends to infinity. The critical exponents associated to the simple random walk are reflected in its Brownian scaling limit both through the exponents in the distributional scaling law stating that $(B_t)_{t\geq 0}$ has the same distribution as $(\lambda^{-1/2}B_{\lambda t})_{t\geq 0}$ and through the fact that $\{B_t:t\geq 0\}$ has both Hausdorff and Minkowski dimension $2$ almost surely when $d\geq 2$. From the point of view of statistical mechanics this foundational example is sometimes described as ``trivial'' since it is ``non-interacting'', and a central goal of the field is to understand critical exponents and scaling limits in models that include some kind of non-trivial interaction. (Of course, the various ``trivial'' models we discuss here
 are all very interesting mathematically and have deep theory associated to their continuum limits.)

An important (but still mostly conjectural) feature of critical exponents and scaling limits is \emph{universality}: the phenomenon by which the large-scale behaviour of a critical system is determined by a small number of qualitative properties of the system (such as its symmetries) and is often independent of small-scale details such as the exact choice of underlying lattice structure. 
Systems with the same large-scale critical behaviour in this sense are said to belong to the same \emph{universality class}. 
Empirically, such universality phenomena are sufficiently strong that entirely unrelated physical systems (meaning real systems that can be studied experimentally in a laboratory!) may exhibit \emph{the same} critical behaviour as each other and as highly idealized mathematical models. Indeed, the \emph{Ising model} can be used to accurately predict the critical behaviour of a variety of unrelated physical systems including gas-liquid transitions, magnetic Curie points, and order—disorder
transitions in alloys~\cite{brush1967history}.
 The universality phenomenon is also related to the phenomenon of \emph{emergent symmetry}, where e.g.\ critical systems defined on square grids that are only invariant under 90 degree rotations at the discrete level may have large-scale limits that have full rotation invariance or even conformal invariance \cite{duminil2020rotational,smirnov2001critical}. (In contrast, \emph{off-critical} systems often retain visible lattice effects at macroscopic scales; search online for images of cubic pyrite crystals for a particularly striking real-world example!)
For example, Brownian motion arises not just as the scaling limit of simple random walk on $\Z^d$ but of any mean-zero random walk on $\R^d$ with finite variance (as well as many processes with \emph{weakly dependent} steps)
and enjoys full rotational symmetry even though the simple random walk on $\Z^d$ is only invariant under 90 degree rotations. (This is an extension to \emph{processes} of the most foundational universality principle in probability theory: the central limit theorem.)
 In addition to its physical significance, the universality phenomenon also lends partial justification to the mathematician's preference to study critical phenomena primarily through a handful of ``canonical'' mathematical models such as the Ising and $\varphi^4$ models, percolation, self-avoiding walk, the Potts model, and so on. On the other hand, for systems with non-trivial interactions, \emph{proofs} of universality and emergent symmetry have mostly remained elusive even for these idealized models, even in two dimensions where the continuum theory is now rather mature.

 \begin{figure}[t]
\centering
\includegraphics[width=0.47\textwidth]{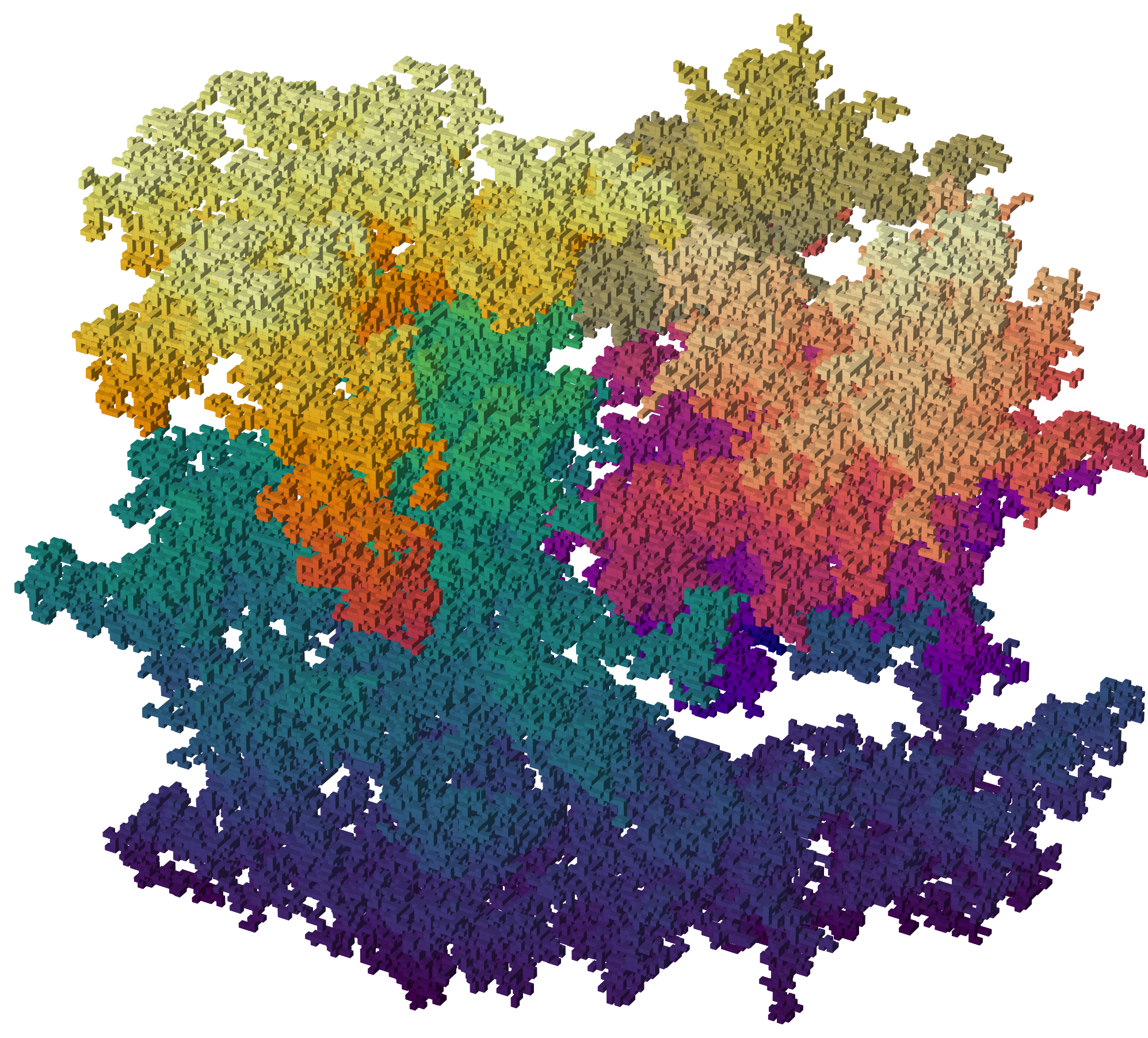}
\hspace{0.5cm}
\includegraphics[clip, trim = {1cm 1cm 1cm 0cm}, width=0.37\textwidth]{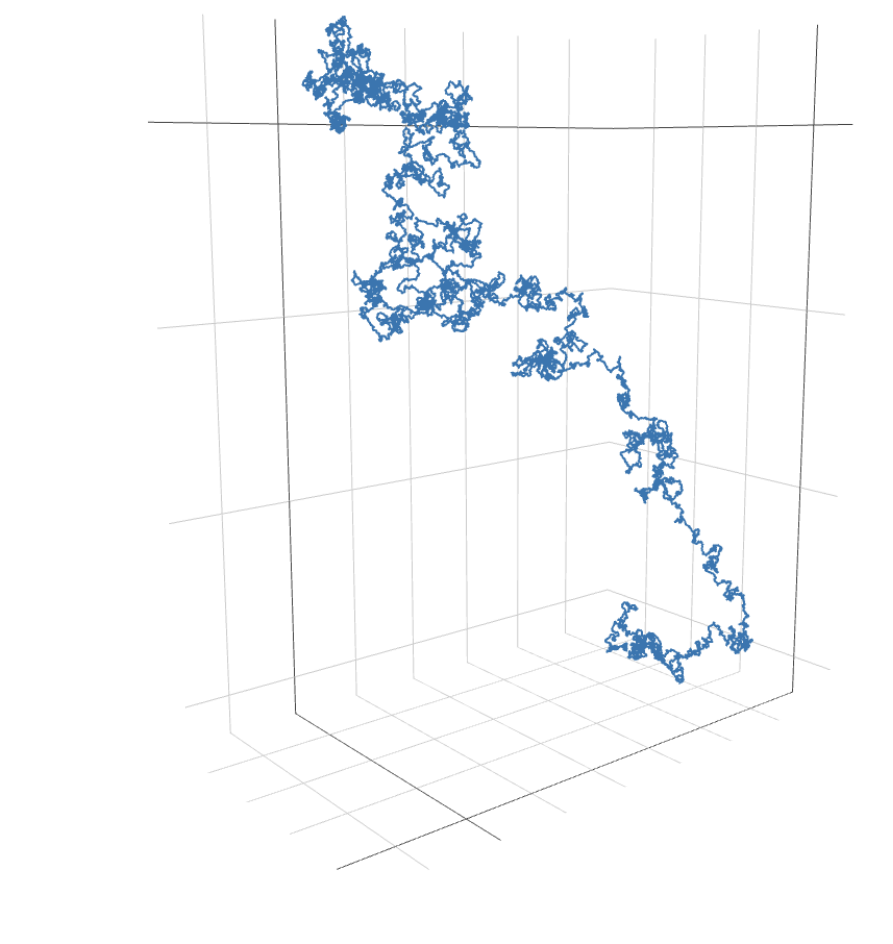}
\caption{Three-dimensional critical phenomena. Left: The five largest clusters in critical site percolation on a three-dimensional box of side-length 2000. Right: A three-dimensional self-avoiding walk with 100,000 steps generated by Ben Wallace ({\color{blue}\url{github.com/bencwallace/polymers.jl}}). Numerically, critical 3d percolation clusters have fractal dimension around $2.52\ldots$ \cite{xu2014simultaneous} while 3d SAWs have fractal dimension around $1.70\ldots$ \cite{clisby2010accurate}.}
\end{figure}

Although the precise choice of $d$-dimensional lattice on which our model is defined should not usually affect its critical behaviour, the choice of dimension $d$ is very important.
 In particular, many models are either known or predicted to have an  \emph{upper critical dimension} $d_c$ such that if $d>d_c$ then the model has relatively simple, ``mean-field'' critical behaviour, while if $d<d_c$ then the model has more complex critical behaviour with both quantitative and qualitative distinctions from the mean-field regime. At the critical dimension $d=d_c$ itself, one typically expects polylogarithmic corrections to mean-field critical behaviour as we discuss in more detail below. The transition to mean-field critical behaviour in high dimensions is often explained in terms of critical models in sufficiently high dimensions behaving as they would in ``geometrically trivial'' settings such as the $3$-regular tree or the complete graph. For models that can be thought of in terms of ``adding an interaction'' to a ``trivial'' reference model, it is often better to think of mean-field critical behaviour as meaning that this interaction becomes \emph{irrelevant} at large scales, with the interacting model having the same large-scale critical behaviour as the reference model in sufficiently high dimensions.

 This is perhaps easiest to understand in the context of \emph{self-avoiding walk} (SAW), i.e., simple random walk (SRW) conditioned not to visit any vertex more than once. In the above idiom, we think of SAW as formed by ``adding an interaction'' to SRW, which is now our ``trivial reference model''. Although the probability that an $n$-step SRW is self-avoiding is always exponentially small in $n$ regardless of the dimension, the nature of the conditioning is (conjecturally) very different according to whether the dimension is bigger, smaller, or equal to the critical dimension $d_c=4$. This can be understood heuristically in terms of a theorem of Erd\H{o}s and Taylor~\cite{ErdTay60} stating that $4$ is also the critical dimension for two independent SRWs to intersect infinitely often. Thus, the self-intersections of a SRW are ``localized'' when $d>4$ but ``macroscopic'' when $d<4$, and in fact Brownian motion in $\R^d$ is almost surely a simple curve if and only if $d\geq 4$. (In 4d, two independent SRWs have polylogarithmically many intersections up to time $n$ and no intersections on each particular large dyadic annulus with high probability, so that these intersections vanish in the scaling limit \cite{lawler2012intersections}.)
This supports the conjecture that SAW and SRW have the same large-scale behaviour when $d>4$ but not when $d<4$. 

The high-dimensional part of this conjecture is now well-understood following the milestone works of Brydges and Spencer \cite{MR782962} (who studied \emph{weakly self-avoiding walk}) and Hara and Slade \cite{MR1171762} (who extended the Brydges-Spencer analysis to \emph{strictly} self-avoiding walk). These works introduced and brought to maturity a technique known as the \emph{lace expansion}, which has now been applied systematically to prove mean-field critical behaviour in a wide range of high-dimensional statistical mechanics models including percolation \cite{MR1043524}, the Ising model \cite{sakai2007lace}, and lattice trees and animals \cite{derbez1997lattice}; see \cite{MR2239599} for a survey.
Following these works, it is now known in particular that SAW on $\Z^d$ has Brownian motion as its scaling limit when $d>4$ \cite{slade1989scaling}, meaning that for each $d>4$ there exists a constant $C$ such that if $(X_m)_{m=0}^n$ is an $n$-step SAW on $\Z^d$ then
\begin{equation}
  \bigl(C n^{-1/2} X_{\lfloor tn \rfloor}\bigr)_{t=0}^1 \longrightarrow (B_t)_{t=0}^1
  \label{eq:SAW_to_BM}
\end{equation}
in distribution as $n\to\infty$.  In the critical dimension $d=d_c=4$ itself it is conjectured that SAW still converges to Brownian motion once appropriate logarithmic corrections to scaling are taken into account, meaning more concretely that the scaling factor $Cn^{-1/2}$ in \eqref{eq:SAW_to_BM} should be replaced by $C n^{-1/2}(\log n)^{-1/8}$. The precise power of the logarithm appearing here has been computed using non-rigorous \emph{renormalization group} (RG) methods \cite{larkin1996phase}, which we discuss in (slightly) more detail below.
 For SAW this conjecture remains open despite significant progress on related problems for the four-dimensional \emph{continuous time weakly self-avoiding walk} via rigorous  RG methods \cite{bauerschmidt2015critical,MR3339164,bauerschmidt2017finite}. The predicted convergence of four-dimensional SAW to Brownian motion at the critical dimension once appropriate polylogarithmic corrections to scaling are taken into account is characteristic of ``marginal irrelevance'' of the self-avoidance interaction (a.k.a.\ ``marginal triviality'' of SAW) at the critical dimension, and similar phenomena are either predicted or proven to occur in a wide range of other models \cite{aizenman2019marginal,bauerschmidt2020three,essam1978percolation,hutchcroft2023logarithmic,lawler2020logarithmic} (see also \cite{fernandez2013random}). Let us also mention that logarithmic corrections to scaling at the critical dimension are often indicative of \emph{very slow} convergence of the discrete model to its continuum limit (with second-order corrections to leading order asymptotic of moments etc.\ only smaller than leading order by a polylogarithmic term), which can make the numerical study of these models via computer simulations particularly difficult at and near the critical dimension.
Below the upper critical dimension the self-interaction is predicted to be sufficiently strong that it completely changes the large-scale behaviour of the model, leading to different critical exponents and scaling limits. 
Very little is known about this rigorously in the context of SAW (although there are very precise \emph{conjectures} in two dimensions \cite{lawler2002scaling,nienhuis1982exact}), and we will defer our detailed discussion of low-dimensional critical phenomena until our discussion of percolation in the next section. 



\section{Critical exponents and scaling limits for Bernoulli percolation: The dream.}
\label{sec:critical_exponents_and_scaling_limits_for_bernoulli_percolation_the_dream_}

Let us now see how this story applies in the case of \emph{Bernoulli bond percolation}, which is the main focus of this article. This model is defined by setting each edge of the hypercubic lattice $\Z^d$ to be either \emph{open} (retained) or \emph{closed} (deleted) independently at random, with each edge having probability $p\in [0,1]$ of being open. 
Percolation was introduced by Broadbent and Hammersley in the 1950s to study porous media \cite{MR91567} and has also been applied to study phase transitions in the formation of gels \cite{coniglio1979site} and the spread of disease \cite{meyers2007contact}. Mathematically, percolation is distinguished by its very simple definition, which often means we can find percolation processes embedded inside other more complex systems and apply percolation methods to study these systems. A comprehensive treatment of the basic theory of percolation can be found in \cite{grimmett2010percolation}.

We write $\P_p$ and $\E_p$ for probabilities and expectations taken with respect to the law of the configuration on $\Z^d$ with parameter $p$. We are interested primarily in the geometry of \emph{clusters}, i.e., connected components of the open subgraph, and define the \emph{critical probability} $p_c=p_c(\Z^d)$ by
$p_c=\inf\{p\in [0,1]: \text{An infinite cluster exists $\P_p$-a.s.}\}$.
(Note that the existence of an infinite cluster is a tail event for the independent Bernoulli random variables defining the model and therefore has probability $0$ or $1$ by Kolmogorov's $0$-$1$ law.) It is a classical theorem (first proven for the Ising model by Peierls \cite{peierls1936ising}) that percolation in dimension $d>1$ has a non-trivial phase transition in the sense that $0<p_c<1$. It is conjectured that there are no infinite clusters at the critical point in every dimension $d\geq 2$; this was proven for $d=2$ by Harris and Kesten \cite{harris1960lower,kesten1980critical} and by Hara and Slade~\cite{MR1043524} in high dimensions (what this means will be made more precise below) but remains a notorious open problem in dimensions $d=3,4,5,6$.

The focus of this survey is not on this qualitative question but instead on the question that comes after it, namely the \emph{quantitative} understanding of critical phenomena in percolation. When $p=p_c$ we expect that all clusters are finite a.s.\ but that large finite clusters are reasonably common, with fractal-like macroscopic clusters existing on each large scale and with the volume and diameter of the cluster of the origin having power law tails. (For $p<p_c$ the volume of the cluster of the origin has an \emph{exponential tail}, a fact of foundational importance known as the \emph{sharpness of the phase transition} \cite{aizenman1987sharpness,duminil2015new,MR852458}.) Some of the most important quantities to understand include the \emph{two-point function} $\P_{p}(x\leftrightarrow y)$, the \emph{susceptibility} $\chi(p):=\E_p|K|=\sum_{x\in \Z^d} \P_{p}(0\leftrightarrow x)$, the \emph{volume tail} $\P_p(|K| \geq n)$, and the \emph{one-arm probability} $\P_p(0\leftrightarrow \partial [-r,r])$, all of which are conjectured to have power-law scaling at and/or near criticality as summarized in \cref{tab:exponents}. 

Just as self-avoiding walk can be understood as a self-interacting version of simple random walk, percolation can be understood as a self-interacting version of \emph{branching random walk}. Here, branching random walk (BRW) on $\Z^d$ is a Markov process describing an evolving cloud of finitely many particles on $\Z^d$, where at each time step every particle splits independently into a random (possibly zero) number of offspring particles, each of which is located at a random neighbour of their parent. If we forget the locations of the particles and observe only their number we get a \emph{branching process}, and the classical theory of such processes yields a phase transition between an extinction phase and a survival phase as the mean offspring per particle exceeds $1$. 
The critical behaviour of branching processes is a classical problem \cite{slack1968branching}, and in the finite-variance case it is known for example that the total progeny (i.e., number of particles that exist at any time) exceeds $n$ with probability of order $n^{-1/2}$, that the probability the process survives for $t$ generations is of order $t^{-1}$, and that the expected total progeny of a mean $1-\eps$ branching process is of order $\eps^{-1}$. Moreover, since the expected number of particles alive in generation $t$ is equal to the $t$th power of the mean offspring number and hence constantly equal to $1$ at criticality, it follows by linearity of expectation that the critical BRW two-point function (i.e., the expected number of particles that visit $y$ when we start the process with a single particle at $x$) is exactly equal to the SRW Green's function (i.e., the expected number of times the random walk started at $x$ visits $y$), which is denoted $G(x,y)$ and is of order $\|x-y\|^{-d+2}$ when $d>2$ and $x\neq y$. 

As with simple random walk, there is a well-developed theory of the \emph{scaling limit} of critical branching random walk \cite{perkins2002part}. (A key difference with simple random walk is that we must now condition the process to survive for a long time or start with a large number of particles to get a non-trivial limit.)
This scaling limit is known as \emph{super-Brownian motion} and is a \emph{measure-valued diffusion}, that is, a stochastic process taking values in the space of finite measures on $\R^d$ that evolves via the infinitesimal branching and diffusion of the infinitesimal ``particles'' making up the measure.  Alternatively, this scaling limit can be constructed by first taking the scaling limit of the genealogical tree to obtain a random fractal tree known as the \emph{continuum random tree} \cite{aldous1991continuum} and then embedding this continuum random tree into space using an appropriate tree-indexed Brownian motion \cite{le1999spatial}. Similarly to how we thought of SRW as a ``two-dimensional random object'', critical BRW should be thought of as a ``four-dimensional random object'' in the sense that the total number of particles that ever exist in the box $[-r,r]^d$ in a critical BRW conditioned to survive for a long time is of order $r^4$ with high probability and the support of the integrated super-Brownian motion in $\R^d$ has Hausdorff dimension $4$ almost surely when $d\geq 4$. See \cite{slade2002scaling} for a survey of applications of super-Brownian motion to high-dimensional statistical mechanics.


Why should percolation be viewed as an interacting version of BRW? When running a computer simulation of percolation, it is natural to sample the cluster of the origin via a breadth-first search: we first query the status of edges incident to the origin, revealing the points at unit ``chemical distance'' (i.e., graph distance in the cluster)  from the origin, then query the status of edges incident to these vertices that have not yet been revealed, and so on. At each stage of the algorithm, we have a cloud of ``active'' vertices (the chemical distance sphere of some radius from the origin), which leads to a new cloud of active vertices at the next step via independent coin flips for each edge in the boundary of the active vertices that has not already been revealed at an earlier stage of the algorithm. If we ignore the constraint that we cannot reuse edges or vertices we have seen before and the fact that there can be at most one particle per site, we would sample a branching random walk rather than percolation. This observation leads to a \emph{monotone coupling} between percolation and BRW (albeit one that does not map the critical point to the critical point and hence has limited use for the study of critical percolation) and allows us to think of percolation as a ``self-interacting BRW'' analogously to how we thought of SAW as a ``self-interacting SRW''. It turns out that for percolation this self-interaction becomes irrelevant at large scales above an upper critical dimension of $d_c=6$, although the heuristic computation of this critical dimension is more subtle than for SAW and will be returned to later. Let us note that percolation is far from the only important model that can be viewed as a self-interacting BRW; other such models include the contact process, oriented percolation, and branched polymers, which have upper critical dimensions $2$, $4+1$, and $8$ respectively \cite{derbez1997lattice,durrett1999rescaled,van2003convergence}.

 In their milestone work \cite{MR1043524}, Hara and Slade used the lace expansion to prove that high-dimensional percolation satisfies the \emph{triangle condition}, a sufficient condition for mean-field critical behaviour introduced by Aizenman and Newman \cite{MR762034}. Following this and other important works including \cite{MR1127713,MR2393990,MR1959796,hutchcroft2023high,MR2551766,MR2748397}, it is now known that high-dimensional percolation satisfies the mean-field estimates
\[
  \P_{p_c}(|K|\geq n) \asymp n^{-1/2}, \qquad \E_{p_c-\eps}|K| \asymp \eps^{-1}, \qquad \text{and} \qquad \P_{p_c}(x\leftrightarrow y) \asymp \|x-y\|^{-d+2},
\]
so that the order of each quantity coincides with that of its BRW analogue. Moreover, following substantial partial progress made in \cite{MR1773141,MR1757958} and the independent parallel work \cite{chatterjee2025limiting}, the full super-Brownian scaling limit of the model is established in the forthcoming work \cite{HutBR_superprocesses}.
An important caveat to these results is that the lace expansion is a \emph{perturbative} method, meaning that it needs a ``small parameter'' to work and cannot be used to prove mean-field critical behaviour under the minimal assumption that $d>d_c=6$. As such, one must either replace the hypercubic lattice with a ``spread-out'' lattice in which two vertices are considered adjacent if they have distance at most some large constant $L$ (in which case any $d>6$ works) or work with the standard hypercubic lattice but take $d\gg 6$ (Hara and Slade needed $d\geq 19$; it is now known that $d\geq 11$ suffices \cite{fitzner2015nearest}). In fact the same caveats apply to SAW but were not mentioned previously since Hara and Slade were able to optimize their arguments well enough to apply to the standard hypercubic lattice in every dimension $d\geq 5$ \cite{MR1171762}. Although there are now other approaches to high-dimensional critical percolation and SAW that do not use the lace expansion \cite{duminil2024alternative,duminil2024alternativeSAW}, these approaches have similar drawbacks to the lace expansion and it remains a major open problem to prove mean-field critical behaviour for high-dimensional percolation and SAW under the minimal assumption that $d>d_c$.

\begin{table}[t] 
\centering

\begin{tabular}{ll lr c c c} 
\toprule
\multicolumn{2}{c}{Quantity} & \multicolumn{2}{c}{Behaviour}  & \multicolumn{1}{c}{Exponent}  & \multicolumn{1}{c}{HD}& \multicolumn{1}{c}{2d} \\
\midrule

Percolation probability & $\theta(p) = \P_p(|K| = \infty)$ & $ \approx (p - p_c)^\beta$ &\text{as $p\downarrow p_c$} & $\beta$ & $1$ & $5/36$\\
\addlinespace
Susceptibility & $\chi(p) = \E_p|K|$ & $\approx |p - p_c|^{-\gamma}$ & \text{as $p\uparrow p_c$}& $\gamma$ & $1$ & \;$43/18$\;\\
 \addlinespace
Cluster volume tail & $\P_{p_c}(|K| \geq n)$ &$\approx n^{-1/\delta}$ & \text{as $n\uparrow \infty$} & $\delta$ & $2$ & $91/5$\\
\addlinespace
Two-point function & $\P_{p_c}(0 \leftrightarrow x)$ &$\approx \|x\|^{-d+2-\eta}$ & \text{as $\|x\|\uparrow \infty$} & $\eta$ & $0$ & $5/24$\\
\addlinespace
Correlation length & $\xi(p)$ & $\approx |p - p_c|^{-\nu}$ & \text{as $p\uparrow p_c$}& $\nu$ & $1/2$ & $4/3$\\
\addlinespace
One-arm probability & $\P_{p_c}(0\leftrightarrow \partial [-r,r]^d)$& $\approx r^{-1/\rho}$ &\text{as $r\uparrow \infty$}& $\rho$ & $1/2$ & $48/5$ \\
\addlinespace
Characteristic volume & $\zeta(p)$
 & $\approx |p - p_c|^{-\Delta}$ & \text{as $p\uparrow p_c$}& $\Delta$ & $2$ & $91/36$\\
\addlinespace
 & $\zeta(p_c,r)$
  & $\approx r^{d_f}$ & \text{as $r\uparrow \infty$}& $d_f$ & $4$ & $91/48$\\
\bottomrule
\end{tabular}
\caption{Some of the most important critical exponents for nearest-neighbour percolation and their  values in high dimensions and in two dimensions. We write $K$ for the cluster of the origin. 
 }
\label{tab:exponents}
\end{table}

 Below the upper-critical dimension it is predicted that the interaction remains relevant at large scales, so that percolation has different critical exponents and scaling limits to BRW. 
Among dimensions $d<d_c$, the two dimensional case is special for a variety of reasons (planar duality, the restricted topology of plane curves, the relevance of complex analysis and the Riemannan mapping theorem, and so on) and has been the subject of vastly more progress than the other low-dimensional cases $d=3,4,5$.
Indeed, the only low-dimensional setting in which critical nearest-neighbour percolation is fully understood is \emph{site percolation on the triangular lattice} \cite{lawler2011values,smirnov2010conformal,smirnov2001criticalexponents}, which is proven to have a conformally invariant scaling limit described in terms of \emph{Schramm-Loewner Evolution} (SLE) with parameter $\kappa=6$ and to have rational critical exponents equal to the values predicted by in the non-rigorous work of Nienhuis \cite{nienhuis1982exact} as summarised in \cref{tab:exponents}. Although the same scaling limits and critical exponents are expected for percolation on all planar lattices, such a universality theorem remains a major open problem \cite{duminil2020rotational}. Detailed surveys of the two-dimensional theory can be found in e.g.\ \cite{werner2007lectures}. 

In six dimensions it is conjectured that critical percolation is ``marginally trivial'' in the sense that it still converges to super-Brownian motion once appropriate logarithmic corrections to scaling are taken into account. Indeed, non-rigorous RG analysis \cite{essam1978percolation,ruiz1998logarithmic} has also led to precise conjectures for what these logarithmic corrections are, including e.g.
\vspace{-0.05cm}
\begin{equation}
\label{eq:Z6_predictions}
  \P_{p_c}(|K|\geq n) \asymp \frac{(\log n)^{2/7}}{\sqrt{n}} 
  \quad \text{ and } \quad 
  \P_{p_c}(x\leftrightarrow y) \asymp \|x-y\|^{-d+2} (\log \|x-y\|)^{1/21}.
\end{equation}
In addition to the substantial technical challenges that plague rigorous RG analyses under the best of circumstances, a major obstacle to proving these conjectures is that none of the non-rigorous works analyze percolation directly, but instead work with the $n$-component $\varphi^3$ model or $q$-state Potts model before taking formal $n\to 0$ or $q\to 1$ limits at the end of the analysis. Indeed, although similar obstacles have been overcome for \emph{continuous time weakly self-avoiding walk}, for which rigorous relations with a \emph{supersymmetric} version of the $\phi^4$ model underpin all the rigorous RG analysis at the critical dimension  \cite{bauerschmidt2015critical,MR3339164,bauerschmidt2017finite}, no comparable isomorphism theorem has been found for percolation. This has prevented an analysis of six-dimensional percolation in the same vein as the works \cite{bauerschmidt2015critical,MR3339164,bauerschmidt2017finite}, whose approach to RG is tailored to ``field theories'' such as the $\varphi^4$ model that can be written as a ``Gaussian free field plus weak interaction''.

What can be said about critical exponents in dimensions $d=3,4,5$?
 Unlike in two dimensions, there are no conjectures regarding exact values of critical exponents nor any apparent reason that the exponents should be rational.  Following the seminal work of Wilson and Fisher \cite{wilson1972critical}, the traditional RG approach to analyzing low-dimensional critical phenomena has been via the \emph{$\eps$-expansion}, meaning that instead of analyzing critical phenomena directly in, say, three dimensions, one instead works in ``$d_c-\eps$ dimensions'' and attempts to expand critical exponents and other quantities of interest in powers of $\eps$. For percolation, non-rigorous RG methods (see e.g.\ \cite{gracey2015four}) have yielded predictions including
\begin{equation}
\label{eq:eps_expansion}
  \eta(6-\eps) = \hspace{0.5em}-\frac{1}{21}\eps - \frac{206}{9261} \eps^2 \pm \cdots \qquad \text{and} \qquad
  \delta(6-\eps) = 2 + \frac{2}{7} \eps + \frac{565}{6174} \eps^2 \pm \cdots.
\end{equation}
(These expansions are predicted to yield \emph{divergent} asymptotic series with very complicated coefficients on the large terms.)
Aside from the problem of interpreting what this means, 
the rigorous implementation of such an analysis for percolation has been held back by the same issues preventing a rigorous RG analysis of logarithmic corrections in six-dimensional percolation, and it seems that the two problems should be closely intertwined.
(Indeed, it is not a coincidence that the coefficients of $\eps$ in \eqref{eq:eps_expansion} coincide with the powers of log in \eqref{eq:Z6_predictions}.)
We remark that in theoretical physics there is now another approach to three-dimensional critical phenomena via the so-called \emph{higher-dimensional conformal bootstrap}, which has yielded the most precise predictions to date regarding the three-dimensional Ising model \cite{poland2019conformal}.
The development of these ideas within probability theory remains in an embryonic state despite enormous recent progress on the \emph{two-dimensional} conformal bootstrap \cite{guillarmou2024review}, and is likely to serve as a major challenge for mathematicians for decades to come \cite{MR3874867}.


\textbf{Why $6$?} We now give two different heuristic derivations of the upper critical dimension $d_c=6$ for percolation, each of which will introduce important perspectives on the model. To establish mean-field critical behaviour, an important step is to establish the mean-field asymptotics of the susceptibility $\chi(p):=\E_p|K| \asymp |p_c-p|^{-1}$ as $p\uparrow p_c$, where we write $\asymp$ to denote an equality that holds to within positive multiplicative constants. We understand the growth of the susceptibility primarily by analyzing its \emph{derivative}, which can be expressed geometrically as 
\begin{equation}
\label{eq:RussoZd}
  \frac{d}{dp}\E_p|K| = \E_p \sum_{x\sim 0} |K||K_x| \mathbbm{1}(0\nleftrightarrow x)
\end{equation}
where $K_x$ denotes the cluster of $x$ and $\{0\nleftrightarrow x\}$ denotes the event that $0$ and $x$ do not belong to the same cluster (this is a special case of Russo's formula \cite[Section 2.4]{grimmett2010percolation}). It is easy to prove an \emph{upper bound} on this derivative by noting that, on the event $\{0\nleftrightarrow x\}$, once the cluster $K$ has been revealed, the cluster $K_x$ is equal in distribution to the cluster of $x$ in percolation on the subgraph of $\Z^d$ in which all edges revealed by $K$ are deleted, and hence clearly has a smaller expectation than it has before the conditioning. In other words, we have the differential inequality
  $\frac{d}{dp}\E_p|K| \leq 2d (\E_p |K|)^2$, which implies by calculus that $\E_p|K| \succeq |p_c-p|^{-1}$ as $p\uparrow p_c$. This is one of several \emph{mean-field lower bounds} for percolation, which state that the critical behaviour in general dimension is always ``at least as severe'' as in high dimensions \cite[Chapter 10]{grimmett2010percolation}. 

  To establish mean-field critical behaviour in high dimensions, one seeks to establish a matching lower bound  on the derivative of the same order:
$\frac{d}{dp}\E_p|K| \geq c (\E_p |K|)^2$. In light of \eqref{eq:RussoZd}, this inequality can be interpreted as meaning that the two distinct clusters $K$ and $K_x$ have a ``bounded interaction'' in high dimensions in the sense that $\E_p|K||K_x|\mathbbm{1}(0\nleftrightarrow x)$ is of the same order as if the two clusters were independent.
Naively, one might guess that $\E_p|K||K_x|\mathbbm{1}(0\nleftrightarrow x) \asymp \E_p|K|\E_p|K_x|$ if and only if two independent clusters have a good probability to remain disjoint when conditioned to be large. In light of the standard heuristics that an independent ``$a$-dimensional random object'' and ``$b$-dimensional random object'' intersect heavily when $a+b>d$ and not when $a+b<d$ and that high-dimensional critical percolation clusters should look like critical BRW and hence be ``four dimensional'', this leads to the (incorrect) guess that the critical dimension is $4+4=8$. (This argument is correct for \emph{lattice trees and animals}, where the interaction really does work this way and the critical dimension is $8$.) A more careful analysis of percolation reveals that the correct criterion for mean-field critical behaviour is not that two large independent clusters are disjoint with good probability, but instead that a single large cluster is disjoint from \emph{the path between two typical points} in an independent large cluster with good probability. In the mean-field regime this path should look like a simple random walk (and hence look ``two-dimensional''), leading to the correct prediction for the critical dimension $d_c=4+2=6$. (The triangle condition for mean-field critical behaviour can be thought of as making rigorous a precise version of this criterion.)

\textbf{Heuristic scaling theory.} Let us now give an alternative heuristic derivation of the upper critical dimension $d=d_c$ via the analysis of the \emph{low-dimensional} regime $d<d_c$. This will be done with the aid of (heuristic) \emph{scaling and hyperscaling theory}, an important topic in its own right. We begin by explaining the \emph{scaling relations} (relations between critical exponents predicted to hold in all dimensions) then turn to the \emph{hyperscaling relations} (which are predicted to hold only for $d\leq d_c$ and are more relevant for the computation of $d_c$). (Here we will discuss scaling theory only for $p\leq p_c$; see \cite[Chapter 9]{grimmett2010percolation} for a more detailed discussion including the supercritical regime.) A central assumption of heuristic scaling theory is that for $p\leq p_c$ there exists a quantity $\xi(p)$ known as the \emph{correlation length} and a function $f$ decaying faster than any power such that
\begin{equation}
  \P_{p}(x\leftrightarrow y) \approx \P_{p_c}(x\leftrightarrow y) f\left(\frac{\|x-y\|}{\xi(p)}\right) \approx \|x-y\|^{-d+2-\eta}f\left(\frac{\|x-y\|}{\xi(p)}\right).
  \label{eq:scaling_assumption1}
\end{equation}
We often interpret $\xi(p)$ as the ``scale on which the model begins to look subcritical''. (We reserve the symbol $\approx$ for vague applications in heuristic arguments.) Positing the existence of the exponents $\gamma,\nu$, and $\eta$ as in \cref{tab:exponents}, this assumed formula allows us to compute
\[
  |p-p_c|^{-\gamma}\approx\E_p|K| \approx \sum_{x\in \Z^d}\|x\|^{-d+2-\eta}f\left(\frac{\|x\|}{\xi(p)}\right) \approx \sum_{x\in [-\xi(p),\xi(p)]^d}\|x\|^{-d+2-\eta} \approx \xi(p)^{2-\eta} \approx |p-p_c|^{-(2-\eta)\nu}
\]
leading to the prediction that $\gamma=(2-\eta)\nu$. Similarly, we can posit the existence of a quantity $\zeta(p,r)$, which we call the \emph{characteristic size of large clusters on scale $r$}, such that
\begin{equation}
\label{eq:scaling_assumption2}
  \P_{p} (|K \cap B_r|\geq n) \approx \P_{p_c} (|K|\geq n) g\left(\frac{n}{\zeta(p,r)}\right)
\end{equation}
for some function $g$ decaying faster than any power. Writing $\zeta(p)=\zeta(p,\infty)$ and positing the existence of the \emph{gap exponent} $\Delta$ and the \emph{fractal dimension} $d_f$ to describe the growth of $\zeta(p)$ and $\zeta(p_c,r)$ as in \cref{tab:exponents}, we can compute as above to obtain
the predictions that $(2-\eta) \delta = d_f (\delta-1)$, $\gamma \delta = \Delta (\delta-1)=\nu d_f (\delta-1)$, and that the higher moments of the cluster volume are described by $\E_{p}|K|^p \approx |p-p_c|^{-\gamma+(p-1) \Delta}$ and $\E_{p_c}|K \cap [-r,r]^d|^p \approx r^{2-\eta+(p-1) d_f}$.

We now turn to the hyperscaling relations. The basic idea behind these relations is that in dimensions $d< d_c$ there should be ``$O(1)$ macroscopic critical clusters on each scale'' that determine all important features of the model. These macroscopic clusters should moreover have size roughly $r^{d_f}$.
Thus, for example, we should have that
\[
  \P_{p_c}(x\leftrightarrow y) \asymp \P_{p_c}(x,y \text{ both in largest cluster on scale $\|x-y\|$}) \approx r^{2(d_f-d)}
\]
leading to the relation $2-\eta=2d_f-d$. 
A further hyperscaling relation posits that the best way for the origin to be in a large cluster is to be in the largest cluster on some scale, leading to the relation 
$(d-d_f)\delta=d_f$. As such, when the hyperscaling relations hold the three exponents $\eta$, $d_f$, and $\delta$ all determine each other.
In dimensions $d>d_c=6$ there is instead a soup of roughly $r^{d-6}$ many large-ish clusters of size roughly $r^4$ that all contribute equally to important quantities on each scale.
At the critical dimension itself we expect that there are polylogarithmically many ``typical large clusters'' on each scale and that the hyperscaling relations remain valid at the level of exponents. As such, we get another heuristic computation of the critical dimension $d_c=6$ by finding that it is the only dimension where the mean-field exponents $\eta=0$ and $d_f=4$ satisfy the hyperscaling relation $2-\eta=2d_f-d$. In two dimensions the scaling and hyperscaling relations were proven by Kesten \cite{MR879034} and played a central role in the eventual computation of the critical exponents for site percolation on the triangular lattice. In dimensions $d=3,4,5$ we have only \emph{conditional} results  \cite{MR1716769}, although there are scaling and hyperscaling  \emph{inequalities} (see e.g.\ \cite{gladkov2024percolation,1901.10363,hutchcroft2020power,MR918404}) that are valid in all dimensions and imply that not all exponents can take their mean-field values when $d<6$.

\section{Long-range models and Sak's prediction.}
\label{sec:long_range_models_}
We now turn our attention to \emph{long-range} models, in which vertices interact not only with their neighbours but with all other vertices, in a manner that decays with the distance between the two vertices. 
In long-range percolation on $\Z^d$, this means more concretely that each pair of vertices $x,y$ are connected by an edge with probability $1-e^{-\beta J(x,y)}$, where $J:\Z^d\times \Z^d\to [0,\infty)$ is a symmetric ($J(x,y)=J(y,x)$), translation-invariant ($J(x,y)=J(0,y-x)$), integrable ($|J|:=\sum_{x\in \Z^d}J(x)<\infty$) kernel and $\beta\geq 0$ is the parameter that is varied to induce a phase transition. We will be most interested in the case of power-law kernels $J(x-y)=\|x-y\|^{-d-\alpha}$ for $\alpha>0$, which has a non-trivial phase transition in the sense that $\beta_c=\inf\{\beta \geq 0:$ an infinite cluster exists$\}$ satisfies $0<\beta_c<\infty$ if and only if $d\geq 2$ or $d=1$ and $\alpha \leq 1$ \cite{newman1986one}.

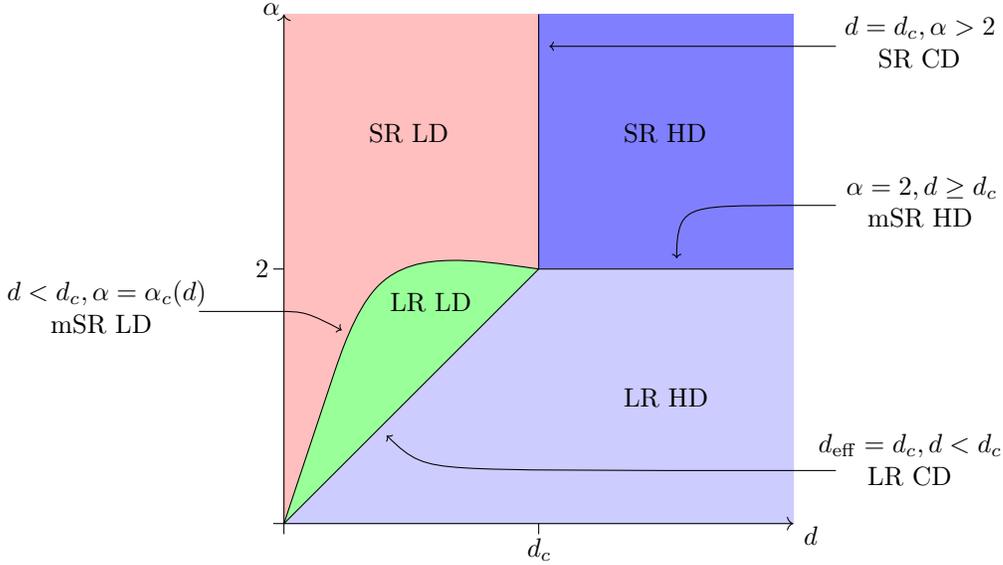
\begin{figure}[t]
\centering
\begin{tikzpicture}[ipe import, baseline, trim left]
  \fill[white]
    (400.3314, 85.8445)
     -- (296.3314, 85.8445);
  \fill[pink]
    (104.3314, 205.8445) rectangle (200.3314, 13.8445);
  \fill[blue!50]
    (200.3314, 205.8445) rectangle (296.3314, 109.8445);
  \filldraw[blue!20]
    (104.3314, 13.8445)
     -- (200.3314, 109.8445)
     -- (296.3314, 109.8445)
     -- (296.3314, 13.8445)
     -- cycle;
  \draw[]
    (200.3314, 109.8445)
     -- (200.3314, 205.8445);
  \draw[]
    (200.3314, 109.8445)
     -- (296.3314, 109.8445);
  \filldraw[, fill=green!40]
    (104.3314, 13.8445)
     .. controls (120.3314, 61.8445) and (120.3314, 61.8445) .. (122.9981, 69.8445)
     .. controls (125.6647, 77.8445) and (130.9981, 93.8445) .. (138.9981, 102.606)
     .. controls (146.9981, 111.3675) and (157.6647, 112.8905) .. (166.9981, 113.0807)
     .. controls (176.3314, 113.271) and (184.3314, 112.1285) .. (200.3314, 109.8445);
  \node[ipe node]
     at (232.331, 57.844) {LR HD};
  \node[ipe node]
     at (232.331, 157.844) {SR HD};
  \node[ipe node]
     at (136.331, 157.844) {SR LD};
  \node[ipe node]
     at (144.331, 93.844) {LR LD};
  \draw[]
    (200.3314, 109.8445)
     -- (104.3314, 13.8445);
  \filldraw[, fill=pink]
    (104.3314, 109.8445)
     -- (100.3314, 109.8445);
  \filldraw[, fill=pink]
    (200.3314, 13.8445)
     -- (200.3314, 9.8445);
  \node[ipe node]
     at (93.659, 106.636) {$2$};
  \node[ipe node]
     at (195.969, 1.49) {$d_c$};
  \node[ipe node]
     at (305.798, 40.814) {$d_\mathrm{eff} = d_c, d<d_c$};
  \draw[ <-]
    (104.3314, 205.8445)
     -- (104.3314, 9.8445);
  \draw[ ->]
    (100.3314, 13.8445)
     -- (296.3314, 13.8445);
  \node[ipe node]
     at (324.331, 28.454) {LR CD};
  \draw[{<[ipe arrow small]}-]
    (252.3314, 113.8445)
     .. controls (252.3314, 117.8445) and (252.3314, 123.8445) .. (254.3314, 127.5111)
     .. controls (256.3314, 131.1778) and (260.3314, 132.5111) .. (267.1314, 133.1778)
     .. controls (273.9314, 133.8445) and (283.5314, 133.8445) .. (312.3314, 133.8445);
  \node[ipe node]
     at (316.331, 137.844) {$\alpha=2, d\geq d_c$};
  \node[ipe node]
     at (324.331, 125.844) {mSR HD};
  \draw[{<[ipe arrow small]}-]
    (204.3314, 193.8445)
     -- (312.3314, 193.8445);
  \node[ipe node]
     at (315.657, 198.101) {$d=d_c, \alpha>2$};
  \node[ipe node]
     at (328.331, 185.844) {SR CD};
  \draw[-{>[ipe arrow small]}]
    (72.3314, 93.8445)
     .. controls (88.3314, 93.8445) and (96.3314, 93.8445) .. (101.6647, 93.8445)
     .. controls (106.9981, 93.8445) and (109.6647, 93.8445) .. (112.9981, 92.8445)
     .. controls (116.3314, 91.8445) and (120.3314, 89.8445) .. (126.1194, 86.7475);
  \node[ipe node]
     at (0, 98.059) {$d<d_c, \alpha=\alpha_c(d)$};
  \node[ipe node]
     at (16.331, 85.844) {mSR LD};
  \node[ipe node]
     at (300.331, 5.844) {$d$};
  \node[ipe node]
     at (96.331, 205.844) {$\alpha$};
  \draw[{<[ipe arrow small]}-]
    (143.0414, 47.0905)
     .. controls (152.3314, 37.8445) and (156.3314, 35.8445) .. (172.3314, 34.8445)
     .. controls (188.3314, 33.8445) and (216.3314, 33.8445) .. (312.3314, 33.8445);
\end{tikzpicture}
\caption{A schematic illustration of the different conjectured regimes of critical behaviour for long-range models including long-range percolation, self-avoiding walk, Ising models, lattice trees, etc. The upper critical dimension $d_c$ is $6$ for percolation, $4$ for self-avoiding walk and the Ising model, and $8$ for lattice trees. The shape of the curve $\alpha=\alpha_c(d)$ pictured here (in which $\alpha_c(d)$ is slightly larger than $2$ for $d$ slightly below $d_c$) is based on numerical analysis of nearest-neighbour percolation together with Sak's prediction $\alpha_c=2-\eta_\mathrm{SR}$; for the Ising model reflection positivity ensures that $\alpha_c(d)\leq 2$ for all $d\geq 1$.}
\label{fig:cartoon}
\end{figure}

There are at least three motivations to consider such systems:
\begin{itemize}
\item Long-range interactions may be justified from a modelling perspective: Physical systems may involve charged particles interacting via inverse-square potentials, diseases may spread not just between neighbours but between any two people living near each other, and so on.
\item Long-range models introduce a new parameter $\alpha$, governing the power-law decay of the interaction kernel, that can sensibly be varied continuously but which has similar effects to varying the dimension. 
This analogy is captured in part by the \emph{effective dimension}
$d_\mathrm{eff}=\max\{d,2d/\alpha\}$,
although we warn the reader that the effective dimension is \emph{not} predicted to fully determine the model's critical behaviour when small.
That $d_\mathrm{eff} >4$ and $d_\mathrm{eff} >6$ are still the correct conditions for mean-field critical behaviour for long-range SAW and percolation  respectively can be justified by the same heuristic calculations as above using that the SRW and BRW with jump kernel proportional to $\|x-y\|^{-d-\alpha}$ converge to an $(\alpha\wedge 2)$-stable L\'evy process and an $(\alpha\wedge 2)$-stable super-L\'evy process with fractal dimensions $(\alpha \wedge 2)$ and $2(\alpha \wedge 2)$ respectively.
The fact that the effective dimension can be varied continuously makes long-range models a natural setting to explore rigorous versions of the $\eps$-expansion, and indeed some aspects of the $\eps$-expansion have rigorously been implemented for the long-range spin $O(n)$ model \cite{MR3723429,MR3772040} (see also \cite{abdesselam2013rigorous}).
\item Perhaps surprisingly, long-range models turn out to be \emph{easier to study} than nearest-neighbour models in some regimes, despite exhibiting a similarly rich variety of critical phenomena. For example, it is a theorem of Berger \cite{MR1896880} that there are no infinite clusters at criticality whenever $\alpha<d$, a regime that is conjectured to include models in the same universality class as nearest-neighbour percolation in each dimension $d\geq 2$. Indeed, the recent work we survey in this article has shown that long-range models can fruitfully serve as a laboratory for the study of critical phenomena beyond the planar and mean-field settings where nearest-neighbour models can be intractably difficult.
\end{itemize}
In addition to all the questions we were interested in for nearest-neighbour models, a further important question for long-range models is to understand when the model is in fact in the same universality class as the nearest-neighbour model. 
More concretely, it is conjectured that long-range percolation with $d>1$ has a \emph{crossover value} $\alpha_c=\alpha_c(d)$ demarcating an \emph{effectively short-range regime} $\alpha>\alpha_c$ (SR) in which the model belongs to the same universality class as the nearest-neighbour model from an \emph{effectively long-range regime} $\alpha<\alpha_c$ (LR) in which the critical behaviour is distinct from that of the short-range model.
Since this transition can occur independently of the transition between low, high, and critical effective dimensions (LD, HD, and CD), and since one expects further subtly distinct critical behaviours in the \emph{marginally short-range} case $\alpha=\alpha_c$ (mSR), this leads to (at least) eight qualitatively different forms of critical behaviour as $d$ and $\alpha$ are varied (\cref{fig:cartoon}).   In one dimension, we instead have that the model is effectively long-range until $\alpha=1$, at which point it is proven to have a \emph{discontinuous phase transition} \cite{MR868738,duminil2020long} and above which there is no phase transition at all.

For SRW/BRW the transition between effectively long-range and short-range regimes always occurs when $\alpha=2$ (the associated SRW having finite variance when $\alpha>2$) as alluded to above, with the walk having a continuous, Brownian scaling limit when $\alpha>2$ and a discontinuous, $\alpha$-stable L\'evy process scaling limit when $\alpha<2$. When $\alpha=2$ the SRW is marginally short-range and still converges to Brownian motion once appropriate logarithmic corrections to scaling are taken into account.

\begin{figure}[t]
\centering
\begin{tikzpicture}[scale=0.5]
\begin{axis}[
    axis lines = left,
    xlabel = $\alpha$,
    ylabel = {$2-\eta$},
    xmin=0,xmax=3,
    ymax=2.5
]
\addplot [
    domain=0:43/24, 
    samples=100, 
    color=red,
    thick
]
{x};
\addplot [
    domain=43/24:3, 
    samples=100, 
    color=red,
    thick
]
{43/24};
\end{axis}

\end{tikzpicture}
    \hspace{0.1cm}
    \begin{tikzpicture}[scale=0.5]
\begin{axis}[
    axis lines = left,
    xlabel = $\alpha$,
    ylabel = {$\delta$},
    xmin=0, xmax=3,
    ymin=0, ymax=25
]
\addplot [
    domain=0:2/3, 
    samples=100, 
    color=red,
    thick
]
{2};
\addplot [
    domain=2/3:43/24, 
    samples=100, 
    color=red,
    thick
]
{(2+x)/(2-x)};
\addplot [
    domain=43/24:3, 
    samples=100, 
    color=red,
    thick
]
{91/5};
\end{axis} 
\end{tikzpicture}
\hspace{0.1cm}
    \begin{tikzpicture}[scale=0.5]
\begin{axis}[
    axis lines = left,
    xlabel = $\alpha$,
    ylabel = {$d_f$},
    xmin=0, xmax=3,
    ymin=0, ymax=2.5
]
\addplot [
    domain=0:2/3, 
    samples=100, 
    color=red,
    thick
]
{2*x};
\addplot [
    domain=2/3:43/24, 
    samples=100, 
    color=red,
    thick
]
{(2+x)/2};
\addplot [
    domain=43/24:3, 
    samples=100, 
    color=red,
    thick
]
{(2+43/24)/2};
\end{axis} 
\end{tikzpicture}
\hspace{0.1cm}
    \caption{Sak's predicted value of $2-\eta$ (left) for $d=2$ and its consequences for $\delta$ (center) and $d_f$ (right) assuming the validity of the hyperscaling relations which give $\delta=(d+\alpha)/(d-\alpha)$ and $d_f=(d+\alpha)/2$ for $d/3<\alpha<\alpha_c$. The predicted crossover value $\alpha_c(2)=43/24$ arises as $2-\eta_{\mathrm{SR}}$ with $\eta_{\mathrm{SR}}=5/24$, $\delta_\mathrm{SR}=91/5$, and $d_{f,\mathrm{SR}}=91/48$.}
    \label{fig:2d}
    \vspace{-1em}
\end{figure}
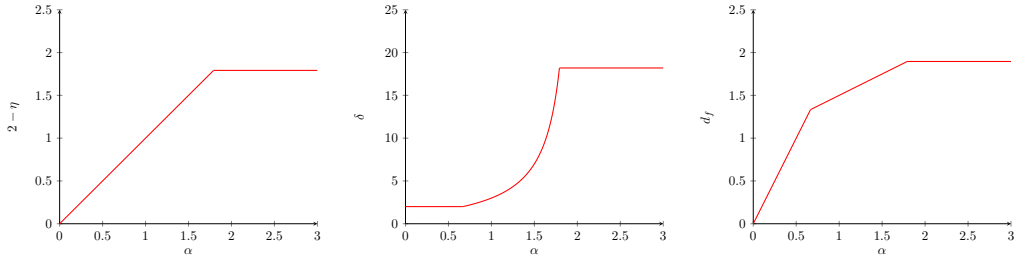

For mathematicians, one of the most appealing features of long-range models is that some (but not all) of their critical exponents are predicted to have simple explicit values throughout the effectively long-range, low-dimensional regime. Indeed, it is a prediction essentially due to Sak \cite{sak1973recursion} (who worked in the context of the $O(n)$ model and built on the earlier works \cite{fisher1972critical,suzuki1972wilson}) that the two-point function critical exponent $\eta$
satisfies
\begin{equation}
\label{eq:Sak_Stick}
2-\eta=\alpha\end{equation}
throughout the entire effectively long-range regime. 
When $d_\mathrm{eff}>d_c$ and $\alpha < 2$ this corresponds to the  two-point function having the same scaling as that of long-range SRW/BRW and has be proven for ``spread-out'' models using the lace expansion \cite{MR2430773,MR3306002}. As such, Sak's prediction is sometimes interpreted as the two-point function ``sticking'' to its mean-field scaling beyond the mean-field regime. It is also predicted that the exponent $\eta$ coincides with its nearest neighbour-value for $\eta_\mathrm{SR}=\eta_\mathrm{SR}(d)$ throughout the effectively short-range regime, so that that
\begin{equation}
\label{eq:Sak}
  2-\eta = \begin{cases} \alpha & \alpha < \alpha_c(d) \qquad \qquad \text{(LR)}
  \\
  2-\eta_{\mathrm{SR}} & \alpha \geq \alpha_c(d) \qquad \qquad \text{(SR and mSR)}
  \end{cases}
\end{equation}
where the crossover value $\alpha_c(d)=2-\eta_\mathrm{SR}$ is the unique value making this function continuous. This prediction is conjectured to hold for a wide range of models, with only the crossover value $\alpha_c(d)=2-\eta_\mathrm{SR}$ being model-dependent. For percolation, the prediction \eqref{eq:Sak} also leads to simple explicit predictions for $d_f$ and $\delta$ via the heuristic scaling and hyperscaling theory discussed above (\cref{fig:2d}). (On the other hand, no closed-form expressions have been conjectured for the exponents $\gamma$, $\nu$, and $\Delta$ in the LR LD regime.)

See \cite{behan2017scaling,brezin2014crossover,gori2017one,luijten1997interaction} for further discussion of Sak's prediction in the physics literature (where it has recently attracted some controversy \cite{liu2025two,picco2012critical}) and \cite{MR3772040,MR3723429} for rigorous partial progress on the spin $O(n)$ model. 

\section{Recent progress on long-range percolation.}
\label{sec:recent_progress_on_long_range_percolation}

We now describe recent work developing a rigorous theory of critical phenomena in long-range percolation. The broad aim of these works is to verify the entire heuristic theory discussed above, including Sak's prediction, the transition between high- and low-dimensional critical behaviour, scaling and hyperscaling theory, superprocess scaling limits in high dimensions, logarithmic corrections to scaling in the critical dimension, and so on, in as many regimes as possible.  So far this has been possible in all regimes other than the effectively short-range low-dimensional regime and its boundaries, and in particular has been possible in the entire effectively long-range regime (including for models that may have low, high, or critical effective dimension). 
For simplicity we will work with the explicit kernel $J(x,y)=\|x-y\|^{-d-\alpha}$, with $\|\cdot\|$ the Euclidean norm, although all the results we state here are universal and apply to a much larger class of kernels than this. 
The works we describe here built on earlier work concerning \emph{hierarchical percolation} \cite{hutchcrofthierarchical,hutchcroft2022critical}, a simplified model of long-range percolation that is always effectively long-range but still exhibits a transition between effectively low-dimensional and low-dimensional regimes with similarly rich critical phenomena to the nearest-neighbour model.

\medskip

\noindent \textbf{The two-point function.} Let us first describe the state of play prior to our recent series of preprints \cite{LRPpaper1,LRPpaper2,LRPpaper3}.
As mentioned above, the lace expansion can be applied when $d_\mathrm{eff}>6$ \cite{MR3306002,MR2430773} (and the doubly-marginal point $d=6$, $\alpha=2$ \cite{MR4032873}) to prove mean-field behaviour of the two-point function under perturbative hypotheses.
Our own first work on the topic \cite{hutchcroft2020power} made Bergers continuity theorem \cite{MR1896880} quantitative, establishing (non-optimal) \emph{power law bounds} on critical long-range percolation with $\alpha<d$.
 In \cite{hutchcrofthierarchical} we showed that the critical two-point function for hierarchical percolation is always of order $\|x-y\|^{-d-\alpha}$, which is consistent with Sak's prediction since hierarchical models are always effectively long-range. In \cite{hutchcroft2022sharp}, we proved that this hierarchical analysis can be extended to the Euclidean model as in the form of a one-sided bound on the \emph{spatial averages} of the two-point function:

\begin{theorem}
\label{thm:two_point_upper}
If $\alpha<d$ then
  $\frac{1}{r^d} \sum_{x\in [-r,r]^d} \P_{\beta_c}(0\leftrightarrow x) \preceq r^{-d+\alpha}$
for every $r\geq 1$. In particular, the critical exponent $\eta$ satisfies $2-\eta\leq \alpha$ if it is well-defined.
\end{theorem}

The proofs of \cite{hutchcroft2022sharp,hutchcrofthierarchical} both rely on what we call a ``runaway observable'' argument: We argue that if certain quantities (such as the median size of the largest cluster on some scale or the expected size of the cluster of the origin) becomes too large on some scale, then it will start growing very rapidly at larger scales. This very rapid growth is impossible for \emph{subcritical} models by sharpness of the phase transition, and a continuity argument shows that it can never be initiated at criticality either. Both arguments also make crucial use of the \emph{universal tightness theorem} \cite{hutchcroft2020power}, which lets us convert bounds on the \emph{median} size of the largest cluster (which is easy to work with in this kind of inductive analysis) to the \emph{moments} of the largest cluster size (which is more useful for what follows). Indeed, this theorem can be thought of as giving a one-sided version of the scaling hypothesis \eqref{eq:scaling_assumption2} with $\zeta(p_c,r)$ replaced by the median size of the largest cluster on scale $r$. 

Besides \cref{thm:two_point_upper} itself, one of the most important takeaways from this proof is that the largest cluster on scale $r$ is always $O(r^{(d+\alpha)/2})$ with high probability at criticality. Note that if $A$ and $B$ are two sets of vertices in the box of radius $r$ then the expected number of edges between them is of order at least $r^{-d-\alpha}|A||B|$, and this bound can be interpreted as saying that the largest critical clusters in two adjacent boxes can never have a high probability to merge by the direct addition of a long edge as we pass from one scale to the next.

Given \cref{thm:two_point_upper}, it is of course very interesting to find conditions under which a matching lower bound holds. This was done by Baumler and Berger in \cite{baumler2022isoperimetric}, who showed that a matching lower bound holds when $\alpha<1$. 
In light of the hyperscaling inequalities of \cite{hutchcroft2020power} (which are also consequences of the universal tightness theorem), this implies that the volume tail exponent $\delta$ satisfies $\delta \geq (d+\alpha)/(d-\alpha)$ when $d\in \{1,2\}$ and $\alpha \in (d/3,1)$, so that the model does \emph{not} have mean-field critical behaviour in this regime.
Building on the techniques of both papers \cite{hutchcroft2022sharp,baumler2022isoperimetric}, it was subsequently proven \cite{hutchcroft2024pointwise} that that these results can be improved to a \emph{pointwise} two-point function bound  under the same assumption that $\alpha<1$. This result completely settles the $d=1$ case of Sak's conjecture for long-range percolation and treats the interval $\alpha\in (2/3,1)$ of non-mean-field models when $d=2$.

\begin{theorem}
\label{thm:two_point_pointwise}
If $\alpha<1$ then $\P_{\beta_c}(x\leftrightarrow y) \asymp \|x-y\|^{-d+\alpha}$ for every $x,y\in \Z^d$. In particular, the critical exponent $\eta$ is well-defined and satisfies $2-\eta=\alpha$ in this regime.
\end{theorem}

The proofs of \cite{baumler2022isoperimetric,hutchcroft2024pointwise} both rely on convolution estimates derived from decomposing the open path between two points according to the first time it exits a \emph{random box}. The threshold $\alpha=1$ arises from a change in the isoperimetry of the pair $(\Z^d,J)$ when $\alpha$ crosses $1$: When $\alpha<1$ the main contribution to the edge boundary of a box $[-r,r]^d$ comes from edges with length of order $r$, whereas for $\alpha>1$ the main contribution comes from edges of bounded length. In light of Sak's prediction, \cref{thm:two_point_pointwise} can be interpreted as establishing the bound $\alpha_c\geq 1$.

\medskip

\noindent \textbf{The real-space RG framework.} It is an interesting feature of Sak's prediction \eqref{eq:Sak} that the critical two-point function does not distinguish between the LD, HD, and CD regimes within the LR regime. In fact the \emph{proofs} of \cref{thm:two_point_upper,thm:two_point_pointwise} are also blind to this distinction, so that different methods are necessary if we want to study more refined critical behaviours that are sensitive to this distinction. To do this, we developed a form of RG analysis for long-range models that can be applied directly to ``geometric'' models of probability theory, without the need to first write the model as a self-interacting Gaussian free field. This was done first for hierarchical models \cite{hutchcroft2022critical} then extended to long-range percolation on $\Z^d$ in \cite{LRPpaper1,LRPpaper2,LRPpaper3}. While this method has the significant drawback that it applies only to long-range models, it has other very considerable advantages:
\begin{itemize}
\item It is technically much easier to implement than classical approaches to rigorous RG. 
\item It can be used to prove \emph{non-perturbative} therorems, without the numerical assumptions needed by the lace expansion or traditional approaches to rigorous RG.
\item It can be used to analyze the entire LR LD regime, in contrast to the traditional RG approach to low-dimensional critical phenomena that require one to work in the $d_c-\eps$ regime.
\end{itemize}
The method also has a much more probabilistic flavour than classical approaches. See \cite[Section I.1.5]{LRPpaper1} for a thorough comparison. Although we will discuss the method only in the context of percolation, it can readily be applied to a variety of other models (at least in the HD regime; the LD and CD analyses may require more work to adapt), and works doing this are already in progress for SAW and the Ising model.

Let us now outline the rudiments of the method, focusing initially on the HD and CD regimes. 
We write $J_r$ for the \textbf{cut-off kernel}
\[
J_r(x,y) := \mathbbm{1}(x\neq y, \|x-y\|\leq r) \int_{\|x-y\|}^r (d+\alpha)s^{-d-\alpha-1} \dif s,
\]
so that $\lim_{r\to \infty}J_r=J$, 
and write $\P_{\beta,r},\E_{\beta,r}$ for probabilities and expectations taken with respect to the law of long-range percolation on $\Z^d$ with kernel $J_r$ at parameter $\beta$. We will always write $\beta_c$ for the critical parameter associated to the original kernel $J$, so that the cut-off measure $\P_{\beta_c,r}$ is subcritical at $\beta_c$ when $r<\infty$. 
At its core, our method aims to compute asymptotics of moments such as $\E_{\beta_c,r}|K|$, $\E_{\beta_c,r}|K|^2$, and $\E_{\beta_c,r}\sum_{x\in K}\|x\|_2^2$ as functions of the cut-off parameter $r$ before using these asymptotics to deduce properties of the original measure $\P_{\beta_c}$ via a Tauberian analysis. In high and critical effective dimension, we do this primarily by finding \emph{asymptotic ODEs} satisfied by these quantities.
Let us first consider the first moment $\E_{\beta_c,r}|K|$. Analogously to \cref{eq:RussoZd}, we can write down the exact derivative formula
\[
  \frac{d}{dr} \E_{\beta_c,r} |K| = (d+\alpha)\beta_c r^{-d-\alpha-1} \E \left[\sum_{\|y\|\leq r} |K||K_y| \mathbbm{1}(0\nleftrightarrow y)\right].
\]
The key difference from the $p$ derivative we considered previously is that most points $y$ contributing to this sum are now very far from the origin when $r$ is large. As such, in high effective dimensions it is plausible that the two clusters $K$ and $K_y$ do not just have a ``bounded interaction'' as before but are in fact \emph{asymptotically independent}, meaning that we can \emph{factor} the expectation appearing here to first order.
Similar expressions can be derived for the derivatives of each moment of the cluster volume which, under the asymptotic independence assumption, lead to the simple asymptotic ODEs
\begin{equation}
\label{eq:volume_moments_ODE_intro}
  \frac{d}{dr}\E_{\beta_c,r}|K|^p \sim  \frac{\pi^{d/2}(d+\alpha) \beta_c}{\Gamma(d/2+1)} r^{-\alpha-1} \sum_{\ell=0}^{p-1}\binom{p}{\ell} \E_{\beta_c,r}|K|^{\ell+1}\E_{\beta_c,r}|K|^{p-\ell}
\end{equation}
as $r\to \infty$, where $\pi^{d/2}/\Gamma(d/2+1)$ is the Lebesgue measure of the unit ball in $\R^d$. In fact we show that these asymptotic ODEs hold not just in the HD regime but also in the LR CD regime, although they are much harder to justify in this latter regime.
More precisely, we prove that the asymptotic ODEs \eqref{eq:volume_moments_ODE_intro} are valid under any of the following conditions, denoted by (HD+):
\begin{itemize}
  \item $d>3\alpha$ or $d=3\alpha<6$.
    \item $d>6$ and  $\P_{\beta_c}(x\leftrightarrow y) \preceq \|x-y\|^{-d+2}$ for all $x\neq y$ 
    \item
     $d=6$, $\alpha=2$, and  $\P_{\beta_c}(x\leftrightarrow y) \preceq (\log \|x-y\|)^{-1}\|x-y\|^{-d+2}$ for all $x\neq y$.
  \end{itemize}
Thus, our results treat the entire regimes LR HD and LR CD, most of the regime mSR HD, and part of the regime SR HD non-perturbatively, without any numerical assumptions on the model, while the rest of the regimes SR HD and mSR HD are treated under the perturbative assumptions needed to implement the lace expansion \cite{MR3306002,MR4032873}. (In this latter case our results still yield significantly stronger results than has been established using the lace expansion alone, including the superprocess scaling limit of the model. In fact \eqref{eq:volume_moments_ODE_intro} is also established non-perturbatively on the entire line $d=3\alpha$, but some of our other conclusions have not yet been established non-perturbatively on this line.)
 Once established, this system of asymptotic ODEs can be solved to yield that
\[
  \E_{\beta_c,r}|K| \sim \frac{\Gamma(d/2+1) \alpha }{\pi^{d/2}(d+\alpha) \beta_c} r^\alpha \quad \text{ and } \quad
  \E_{\beta_c,r}|K|^p \sim (2p-3)!! \left(\frac{\E_{\beta_c,r}|K|^2}{\E_{\beta_c,r}|K|}\right)^{p-1} \E_{\beta_c,r}|K|.
\]
On the other hand, the asymptotic ODEs \eqref{eq:volume_moments_ODE_intro} guarantee only that the second moment $\E_{\beta_c,r}|K|^2$ is regularly varying of index $3\alpha$, so that logarithmic corrections to scaling may be present in this second moment. However, they also guarantee that the second moment is the \emph{only} place where logarithmic corrections may enter the system, with corrections that enter at the second moment replicated in a simple way at higher moments. In the HD regime the errors in the ODE \eqref{eq:volume_moments_ODE_intro} are small enough that the second moment is asymptotic to a constant multiple of $r^{3\alpha}$, whereas in the LR CD regime we must analyze the asymptotics of the \emph{second-order corrections} to the ODE \eqref{eq:volume_moments_ODE_intro} to compute the logarithmic corrections to scaling, with the second moment eventually shown to be asymptotic to a constant multiple of $(\log r)^{-1/2} r^{3\alpha}$. (Interestingly, the computation of these second-order corrections requires us to first know the full superprocess scaling limit of the model, so that we prove this scaling limit result with respect to unknown scaling factors before using our knowledge of the scaling limit to compute these scaling factors.)


This analysis of moments eventually leads to the following theorem via a Tauberian analysis, which provides a bridge from the asymptotic behavior of moments in the regularized system (as the cut-off goes to infinity) to the power-law tail behavior of distributions in the original, unregularized model.

\begin{theorem}[HD and LR CD critical volume tail]
If at least one of the hypotheses of \emph{(HD+)} holds then
\[
  \P_{\beta_c}(|K|\geq n) \sim \mathrm{const.}\; \begin{cases} n^{-1/2} & d_\mathrm{eff}>6 \text{ or } d=6,\alpha=2 \qquad \text{\emph{(HD)}}\\
(\log n)^{1/4} n^{-1/2} & d_\mathrm{eff} =6, \alpha<2 \hspace{2.15cm} \text{\emph{(LR CD)}}
  \end{cases}
\]
as $n\to\infty$.
\end{theorem}

The full scaling limit of the HD/CD model can be understood via the asymptotic analysis of moments of the form $\E_{\beta_c,r}\sum_{x_1,\ldots,x_n\in K}\prod_{i=1}^n P_i(x_i)$ with $P_1,\ldots,P_n$ polynomials. This analysis also reveals the transition between the LR HD and SR HD regimes when $\alpha=2$. For example, we prove under (HD+) that $\E_{\beta_c,r}\sum_{x\in K}\|x\|_2^2$ satisfies an asymptotic ODE of the form
\begin{equation}
\frac{d}{dr} \E_{\beta_c,r} \left[\sum_{x\in K} \|x\|^2 \right]
\sim  \frac{2 \alpha}{r} \E_{\beta_c,r} \left[\sum_{x\in K} \|x\|^2 \right] + C r^{2+\alpha-1},
\label{eq:x12_derivative_asymptotic2_intro}
\end{equation}
as $r\to \infty$ for an explicit positive constant $C$, where the second term encodes the contribution to the distance from the pivotal edge itself. This ODE has a change in behaviour according to whether the coefficient $2\alpha$ is larger or smaller than the exponent $2+\alpha$, which encodes the transition between the LR HD and SR HD regimes at $\alpha=2$. This asymptotic analysis of spatial moments eventually leads to the following scaling limit theorem.

\begin{theorem}[Superprocess limits]
\label{thm:superprocess_main}
Suppose that at least one of the hypotheses of \emph{(HD+)} holds.
  There exist functions $\zeta(R)\to\infty$ and $\eta(R)\to 0$ such that
\[
\frac{1}{\eta(R)}\P_{\beta_c}\left(|K|\geq \lambda \zeta(R),\; \frac{1}{\zeta(R)}\sum_{x \in K} \delta_{x/R}\in \;\cdot\;  \right) \to \mathbb{N}\Bigl(\mu(\R^d)\geq \lambda, \mu\in \cdot\Bigr)
\]
weakly as $R\to \infty$ for each $\lambda>0$, where $\mathbb{N}$ denotes either the canonical measure of the integrated symmetric $\alpha$-stable L\'evy superprocess excursion associated to the L\'evy measure with density $\frac{\alpha}{d+\alpha}\|x\|^{-d-\alpha}$ if $\alpha <2$ or the canonical measure of integrated super-Brownian excursion if $\alpha \geq 2$. Moreover, the normalization factors $\zeta$ and $\eta$ are given asymptotically by
\[
\zeta(R) \sim \mathrm{const.}\;\begin{cases}
 R^{4} \\
 R^{4} (\log R)^{-2} \\
R^{2\alpha} \\
R^{2\alpha} (\log R)^{-1/2} 
\end{cases}
\text{ and } \quad
\eta(R) \sim \mathrm{const.}\;\begin{cases}
 R^{-2} &\; \alpha>2, d>6 \qquad\phantom{\alpha} \text{\emph{(SR HD)}}\\
 R^{-2} \log R &\; \alpha = 2, d\geq 6 \qquad \phantom{\alpha} \text{\emph{(mSR HD)}}\\
R^{-\alpha} &\; \alpha < 2, d > 3\alpha \qquad \text{\emph{(LR HD)}}\\
R^{-\alpha} (\log R)^{1/2}  &\; \alpha < 2, d = 3\alpha \qquad \text{\emph{(LR CD)}}
\end{cases}
\]
as $R\to \infty$.
\end{theorem}

Here the word ``integrated'' means we are considering our superprocesses as random \emph{measures} only (integrated over time), and do not keep track of the internal structure of clusters.

\medskip

\noindent \textbf{Defining the effectively long-range regime.} A major difficulty in the study of the effectively \emph{low-dimensional} regime is that the computation of the crossover value $\alpha_c$ is conjecturally equivalent to the computation of the nearest-neighbour exponent $\eta_\mathrm{SR}=2-\alpha_c$, which appears to be beyond the scope of existing techniques. (Indeed, proving this appears to be a very difficult problem even in two dimensions.) To avoid dealing with this issue, we will instead \emph{define} what it means for our model to be effectively long-range. Our definition is motivated by our analysis of the HD and CD regimes, where we compute the \emph{$L^p$ radius of gyration} (a.k.a.\ correlation length of order $p$) of the cut-off measure $\P_{\beta_c,r}$ to be
\begin{equation}
\label{eq:radius_of_gyration_intro}
\xi_{p}(\beta_c,r):= \left[\frac{\E_{\beta_c,r}\left[\sum_{x\in K} \|x\|^{p} \right]}{\E_{\beta_c,r}|K|}\right]^{1/p} \sim 
C_p\cdot\;\begin{cases}
 r^{\alpha/2} &\; \alpha>2, d>6\phantom{\alpha} \qquad \text{(SR HD)}\\
 r \sqrt{\log r} &\; \alpha = 2, d\geq 6\phantom{\alpha} \qquad \text{(mSR HD)}\\
r &\; \alpha < 2, d \geq 3\alpha \qquad \text{(LR HD \& CD)}.
\end{cases}
\end{equation}
The quantity $\xi_p(\beta_c,r)$ measures the distance between two typical points in a typical large cluster under the measure $\P_{\beta_c,r}$, and can be though of as one precise measure of the correlation length. Thus, we see from \eqref{eq:radius_of_gyration_intro} that the LR regime is characterised by this correlation length having the same order as the cut-off scale $r$, with this correlation length instead being much larger than $r$ in the SR and mSR regimes. 

Why is this the reasonable thing to expect? When the model is effectively long-range, we expect that every open path between two distant vertices $x$ and $y$ in the critical cluster should contain a ``macroscopic edge'', that is, an edge with length of order $\|x-y\|$, with high probability on the event that $x$ and $y$ are connected. Conversely, when the model is \emph{not} effectively long-range, such macroscopic edges should be avoidable so that in the scaling limit of the model we see only continuous paths. Interpreted in terms of the cut-off measures $\P_{\beta_c,r}$, this means that in the effectively long-range regime the probability $\P_{\beta_c,r}(x\leftrightarrow y)$ should be significantly smaller than $\P_{\beta_c}(x\leftrightarrow y)$ when $r$ is of order $\|x-y\|$, whereas in the effectively short-range regime $\P_{\beta_c,r}(x\leftrightarrow y)$ does not begin to become significantly smaller than $\P_{\beta_c}(x\leftrightarrow y)$ until $r$ passes below some scale much smaller than $\|x-y\|$. 

\medskip

In light of this discussion, we \emph{define} the LR regime via the validity of the LR case of the estimate \eqref{eq:radius_of_gyration_intro}:

\begin{defn}[CL]
\label{def:CL}
 We say that the model satisfies the \textbf{correlation length condition for effectively long-range critical behaviour} (CL) if $\beta_c<\infty$ and for each $p>0$ there exists a constant $C_p<\infty$ such that $\xi_p(\beta_c,r) \leq C_p r$ for every $r\geq 1$.
\end{defn}

Our current knowledge of when (CL) holds is described in the following theorem. Note in particular that (CL) is known to hold for the effectively low-dimensional models with $d\in \{1,2\}$ and $\alpha \in (d/3,1)$.

\begin{theorem}[The location of the effectively long-range regime]
\begin{itemize}
  \item \emph{(CL)} holds when $\alpha<1$ and does not hold when $\alpha>d$.
\item If $d\geq 3\alpha$ or at least one hypothesis of \emph{(HD+)} holds then \emph{(CL)} holds if and only if $\alpha<2$.
\end{itemize}
\end{theorem}



In \cite{LRPpaper2} we prove the validity of Sak's prediction throughout the LR regime as defined by (CL)
and that (CL) is characterised by the validity of a scaling estimate for the cut-off two-point function in the spirit of \eqref{eq:scaling_assumption1}:

\begin{theorem}[LR critical two-point functions]
\label{thm:CL_Sak}
If \emph{(CL)} holds then
$\P_{\beta_c}(x\leftrightarrow y)\asymp \|x-y\|^{-d+\alpha}$
for every distinct $x,y\in \Z^d$.  Moreover, \emph{(CL)} holds if and only if there exists a decreasing function $h:(0,\infty)\to (0,1]$ decaying faster than any power of $x$ such that
\begin{equation}
  \P_{\beta_c,r}(x\leftrightarrow y)\preceq \|x-y\|^{-d+\alpha} \,h\Biggl(\frac{\|x-y\|}{r}\Biggr)
  \label{eq:CL_h}
\end{equation}
for every $r\geq 1$ and distinct $x,y\in \Z^d$.
\end{theorem}

It remains an important open problem to find further equivalent characterisations of the effectively long-range regime, ideally leading to ``dichotomy theorems'' separating the effectively long-range and short-range regimes and hence to a more complete resolution of Sak's prediction.

\medskip

\textbf{Further analysis of the effectively long-range regime.} Going beyond the two-point function, our first major result about the LD LR regime rigorously establishes the validity of heuristic hyperscaling theory in this regime and uses this theory to compute the exponents $\delta$ and $d_f$:


\begin{theorem}[LR LD critical cluster volumes]
\label{thm:main_low_dim}
If $d<3\alpha$ and 
\emph{(CL)} holds then 
\[
\P_{\beta_c}(|K|\geq n) \asymp 
n^{-(d-\alpha)/(d+\alpha)} 
\qquad \text{ and } \qquad \E_{\beta_c}|K \cap B_r|^p  \asymp_p 
r^{\alpha+(p-1)\frac{d+\alpha}{2}}
\]
for every $n,r\geq 1$ and integer $p\geq 1$.
 In particular, the exponents $\delta$ and $d_f$ are well-defined and given by $\delta=(d+\alpha)/(d-\alpha)$ and $d_f=(d+\alpha)/2$ respectively.
\end{theorem}

Interestingly, the proof of \cref{thm:main_low_dim} proceeds by applying the \emph{high-dimensional} RG analysis sketched above, but now in the context of a proof by contradiction: Roughly speaking, we show that if (CL) holds and the largest cluster on scale $r$ is much smaller than the previously-established upper bound of $r^{(d+\alpha)/2}$ then the asymptotic ODEs \eqref{eq:volume_moments_ODE_intro} are valid and one can deduce that the moments $\E_{\beta_c,r}|K|$ and $\E_{\beta_c,r}|K|^2$ grow with exponents $\alpha$ and $3\alpha$ as in high dimensions. On the other hand, the universal tightness theorem together with our $O(r^{(d+\alpha)/2})$ bound on the largest cluster yield under (CL) that
$\E_{\beta_c,r}|K|^2 \preceq r^{(d+\alpha)/2}\E_{\beta_c,r}|K|$, which is inconsistent with the previous growth estimates when $d<3\alpha$. The hyperscaling relations are established in part by applying a related ODE analysis to the \emph{truncated moment} $\E_{\beta_c,r}\min\{|K|,m\}$ to show that clusters much smaller than $r^{(d+\alpha)/2}$ contribute negligibly to quantities of interest on scale $r$.

Our techniques are also able to establish up-to-constants estimates on \emph{$k$-point functions} throughout the LR regime as defined by (CL). We state these results only for the three-point function, which already reveal further distinctions between the HD, CD, and LD regimes:

\begin{theorem}[LR three-point function]
Suppose that \emph{(CL)} holds. The critical three-point function is given up-to-constants by
\[
  \P_{\beta_c}(x\leftrightarrow y \leftrightarrow z) \asymp d_{\mathrm{max}}(x,y,z)^{-d+\alpha} \cdot \begin{cases} d_\mathrm{min}(x,y,z)^{-d+2\alpha} & d> 3\alpha \qquad \text{\emph{(LR HD)}}\\ 
 d_\mathrm{min}(x,y,z)^{-d+2\alpha} (\log[1+ d_\mathrm{min}(x,y,z)])^{-1/2}  & d=3\alpha \qquad \text{\emph{(LR \hspace{0.05em}CD)}}\\
d_\mathrm{min}(x,y,z)^{-\frac{d-\alpha}{2}} 
&d<3\alpha \qquad \text{\emph{(LR \hspace{0.15em}LD)}}
  \end{cases}
\]
for every  triple of distinct points $x,y,z\in \Z^d$, 
where $d_{\mathrm{max}}(x,y,z)$ and $d_\mathrm{min}(x,y,z)$ are the maximal and minimal distances between the three points $x,y,z$ respectively.
\end{theorem}

This theorem shows that the \emph{tree graph inequality} of Aizenman and Newman \cite{MR762034} is sharp in the LR HD regime, while a very interesting new three-point function inequality due to Gladkov \cite{gladkov2024percolation} is sharp in the LR LD regime. Both inequalities, which are true for percolation on any weighted graph, become equivalent in the critical dimension in which case they are both wasteful by a polylogarithmic factor.

\medskip

\textbf{Scaling relations for slightly subcritical exponents.} As mentioned above, the exponents $\gamma$, $\Delta$, and $\nu$ are not conjectured to have closed-form expressions in the LR LD regime in contrast to $\eta$, $\delta$, and $d_f$. Still, we are able to prove that these exponents satisfy the scaling relations throughout the LR regime assuming that at least one of them is well-defined. The proof of this theorem verifies not only the relations between exponents stated here but long-range analogues of the scaling assumptions \eqref{eq:scaling_assumption1} and \eqref{eq:scaling_assumption2} used in their heuristic derivation.

\begin{theorem}[LR scaling relations]
\label{thm:scaling_relations}
Suppose that \emph{(CL)} holds. If at least one of the exponents $\gamma$, $\nu$, or $\Delta$ is well-defined then all three exponents are well-defined and related by the scaling relations
$\gamma=(2-\eta)\nu$ and $\Delta = \nu d_f$
where $2-\eta=\alpha$ and $d_f = \min\{(d+\alpha)/2,2\alpha\}$.
\end{theorem}

\textbf{Directions for future research.} The works surveyed here leave many promising directions open for further research. One such direction which appears to already be in reach is to perform a rigorous $\eps$-expansion to first order for the exponents $\gamma$, $\nu$, and $\Delta$ in the LR LD regime, which should also establish the non-triviality of the LR LD regime in every dimension $1\leq d \leq 5$. A more ambitious direction is to establish existence and conformal invariance of the model's scaling limit in the LR LD regime; our results already yield various forms of \emph{tightness} (i.e., existence and non-triviality of subsequential limits), including a striking ``up-to-constants conformal invariance'' estimate on the $k$-point function which can be viewed as a form of ``tightness under M\"obius transformations''. Finally, perhaps the most difficult problem will be to prove the \emph{effectively short-range} part of Sak's prediction, meaning that once the model is no longer effectively long-range it must belong to the same universality class as the nearest-neighbour model. As an intermediate step, it would be very interesting to prove a uniqueness theorem for an appropriate axiomatic characterisation of ``percolation scaling limits'', where the short-range assumption manifests itself as some kind of continuity assumption in the limit; this could be compared with the fact that Brownian motion is the only \emph{continuous} stochastic process with independent, stationary, centered increments.

\section*{Acknowledgments.}
This work was supported by NSF grant DMS-1928930 and a Packard Fellowship for Science and Engineering. We thank Yujin Kim for proofreading a draft.

\bibliographystyle{siamplain}
\bibliography{unimodularthesis}
\end{document}